\documentclass[letter,longauth]{aa} 
\usepackage{natbib}
\usepackage[breaklinks=true]{hyperref}
\usepackage[utf8]{inputenc}
\usepackage[varg]{txfonts}
\usepackage{color}
\usepackage{xcolor}
\usepackage{lipsum}
\usepackage[english]{babel}
\usepackage{fontenc,amssymb}
\usepackage{graphicx}
\usepackage{amsmath}
\usepackage{chemformula}
\usepackage[normalem]{ulem}

\graphicspath{{Figures/}}

\newcommand{\cmsq}{\hbox{cm$^{-2}$}}
\newcommand{\cmcu}{\hbox{cm$^{-3}$}}

\newcommand{\um}{\textmu m}

\begin{document}

\title{PDRs4All
\\ XV.
 CH radical and \ch{H3+} molecular ion in the  irradiated protoplanetary disk d203-506}

\author{
    I.~Schroetter \inst{1} \and 
    O.~Berné \inst{1}\and 
    J.~R.~Goicoechea \inst{2} \and
    J.~H.~Black \inst{3} \and
    O.~Roncero \inst{2} \and
    F.~Alarcon \inst{4} \and
    P.~Amiot \inst{1} \and
    O.~Asvany \inst{5} \and
    C.~Boersma \inst{6} \and
    S.~Brünken \inst{7, 8} \and
    J.~Cami \inst{9, 10, 11} \and
    L.~Coudert \inst{12} \and
    E.~Dartois \inst{12} \and
    A.~Fuente \inst{13} \and
    B.~Gans \inst{12} \and
    A.~Gusdorf \inst{14, 15} \and
    U.~Jacovella \inst{12} \and
    M.~A.~Martin~Drumel \inst{12} \and
    T.~Onaka \inst{16} \and
    E.~Peeters \inst{9, 10, 11} \and
    E.~Roueff \inst{17} \and
    A.~G.~G.~M.~Tielens \inst{18, 19} \and
    M.~Zannese \inst{20}    
        }
\institute{Institut de Recherche en Astrophysique et Plan\'etologie, Universit\'e Toulouse III - Paul Sabatier, CNRS, CNES, 9 Av. du colonel Roche, 31028 Toulouse Cedex 04, France \label{inst1} 
\and
Instituto de F\'{\i}sica Fundamental (CSIC). Calle Serrano 121-123, 28006, Madrid, Spain 
\and
Department of Space, Earth, and Environment, Chalmers University of Technology, Onsala Space Observatory, 43992 Onsala, Sweden.
\and
Dipartimento di Fisica, Università degli Studi di Milano, Via Celoria 16, 20133 Milano, Italy
\and
I. Physikalisches Institut, Universität zu Köln, Zülpicher Str.77, D-50937 Köln, Germany
\and
NASA Ames Research Center, MS 245-6, Moffett Field, CA 94035-1000, USA
\and
HFML-FELIX, Toernooiveld 7, 6525ED Nijmegen, the Netherlands
\and
Institute for Molecules and Materials, Radboud University, Heyendaalseweg 135, 6525AJ  Nijmegen, the Netherlands
\and 
Department of Physics \& Astronomy, The University of Western Ontario, London ON N6A 3K7, Canada
\and
Institute for Earth and Space Exploration, The University of Western Ontario, London ON N6A 3K7, Canada
\and
Carl Sagan Center, Search for ExtraTerrestrial Intelligence  Institute, Mountain View, CA 94043, USA
\and
Institut des Sciences Moléculaires d'Orsay, CNRS, Université Paris-Saclay, 91405 Orsay, France 
\and
Centro de Astrobiología (CAB), CSIC-INTA, Ctra. de Ajalvir, km 4, Torrejón de Ardoz, 28850 Madrid, Spain
\and
Laboratoire de Physique de l’École normale supérieure, ENS, Université PSL, CNRS, Sorbonne Université, Université Paris Cité, 75005 Paris, France
\and
Observatoire de Paris, Université PSL, Sorbonne Université, LERMA, CNRS UMR 8112, 75014 Paris, France 
\and
Department of Astronomy, Graduate School of Science, The University of Tokyo, Tokyo 113-0033, Japan
\and 
LUX, UMR 8262, Observatoire de Paris, PSL Research University,
CNRS, Sorbonne Universités, 92190 Meudon, France
\and 
Astronomy Department, University of Maryland, College Park, MD 20742, USA
\and
Leiden Observatory, Leiden University, P.O. Box 9513, 2300 RA Leiden, The Netherlands
\and
Institut d'Astrophysique Spatiale, Universit\'e Paris-Saclay, Centre National de la Recherche Scientifique, 91405 Orsay, France
}

\label{firstpage}
\abstract
{Most protoplanetary disks experience a phase in which they are subjected to strong ultraviolet radiation from nearby massive stars. This UV radiation can substantially alter their chemistry by producing numerous radicals and molecular ions.
In this Letter we present a detailed analysis of the JWST-NIRSpec spectrum of the d203-506 obtained as part of the PDRs4All Early Release Science program. Using state-of-the-art spectroscopic data, we searched for species using
a multi-molecule fitting tool, \textsc{PAHTATmol}, which we developed for this purpose. 
Based on this analysis, we report the clear detection of ro-vibrational emission of the \ch{CH} radical and the {likely} detection of the \ch{H3+} molecular ion, with estimated abundances of a few times 10$^{-7}$ and approximately 10$^{-8}$, respectively. The presence of \ch{CH}  is predicted by gas-phase models and is well explained by hydrocarbon photochemistry. {Interstellar
} \ch{H3+} is usually formed through reactions of \ch{H2} with \ch{H2+} originating from cosmic ray ionization of  \ch{H2}. However, recent theoretical studies suggest that \ch{H3+} 
also forms through far-UV (FUV)-driven chemistry in strongly irradiated 
\mbox{($G_0$\,$>$10$^3$)}, dense \mbox{($n_{\rm H} >10^{6}$ \cmcu)} gas. 
The latter is favored as an explanation for the presence 
of  hot 
\ch{H3+} ($T_{\rm ex}$\,$\gtrsim$\,1000\,K)    in the outer disk layers of d203-506, coinciding with the emission of \mbox{FUV-pumped} H$_2$ and other
\mbox{photodissociation region (PDR) species}, such as CH$^+$, {\ch{CH3+}}, and OH.
{Our detection of infrared emission from vibrationally excited H$_3^+$ and CH raises questions about their excitation mechanisms and\,} 
underscores that FUV radiation can have a profound impact on the chemistry of planet-forming disks. They also demonstrate the power of JWST to push the limit of the detection of elusive species 
{in protoplanetary disks.}  }

\keywords{Protoplanetary disks --- ISM: abundances --- photon-dominated region (PDR) -- techniques: spectroscopic }

\titlerunning{Chemical inventory of d203-506}  
\maketitle

\section{Introduction}

Ultraviolet ({UV}) radiation from massive stars is ubiquitous in galaxies and plays a fundamental role in the physics and the chemistry of star and planet formation \citep[e.g.,][]{wolfire2022photodissociation}. 
JWST spectroscopic observations of protoplanetary disks irradiated by UV radiation from 
massive stars have revealed that this energy input drives a chemistry dominated by 
{abundant} ions and radicals. 
A key signature of this photochemistry is the methyl cation \ch{CH3+} discovered in the highly irradiated ($G_0 = 2 \times 10^4$) protoplanetary disk d203-506 in Orion  \citep{Berne_ch3p, ChangalaB_ch3p}. Other tracers of these processes include the detection of \mbox{rotationally hot} \ch{OH}  \citep{Zannese_24_oh}, the presence of the reactive ion \ch{CH+}  \citep{BerneO_2024_science, zannese2025}, and the infrared \mbox{fluorescent} 
emission {from electronically excited} {C\,{\sc i}}  \citep{Goicoechea2024}. These species  
reside in the upper disk layers  exposed to UV radiation, while most neutral  species survive in the dense inner  disk regions \citep[e.g.,][]{schroetter_2025_503_504}. 
These findings demonstrate that JWST is an unprecedented spectroscopic tool for detecting and characterizing small ions and radicals {triggered by} UV-driven chemistry. This capability is relevant to a wide range of astrophysical environments, from the starburst galaxies to the atmospheres of forming protoplanets in stellar clusters.
Maximizing the potential of JWST requires tools that rely on state-of-the-art spectroscopic data to identify and characterize the excitation properties of the numerous species present in the infrared. 
In this Letter, we present  \textsc{PAHTATmol}, {a tool} dedicated to this task, and applied to NIRSpec spectra of the irradiated protoplanetary disk d203-506.
This tool supports the detection of two new species in a protoplanetary disk: CH and \ch{H3+}.
We present the observations of d203-506 in Sect.~\ref{section:observations} and the fitting process of its NIRSpec spectrum and the results in Sect.~\ref{section:fitting_main}. 
We  describe and discuss {the origin and properties of} the two species 
in Sect.~\ref{section:ch_main}, and  present our conclusions in Sect.~\ref{section:conclusion}.

\begin{figure*}[t]
\centering
\includegraphics[width=0.80\textwidth]{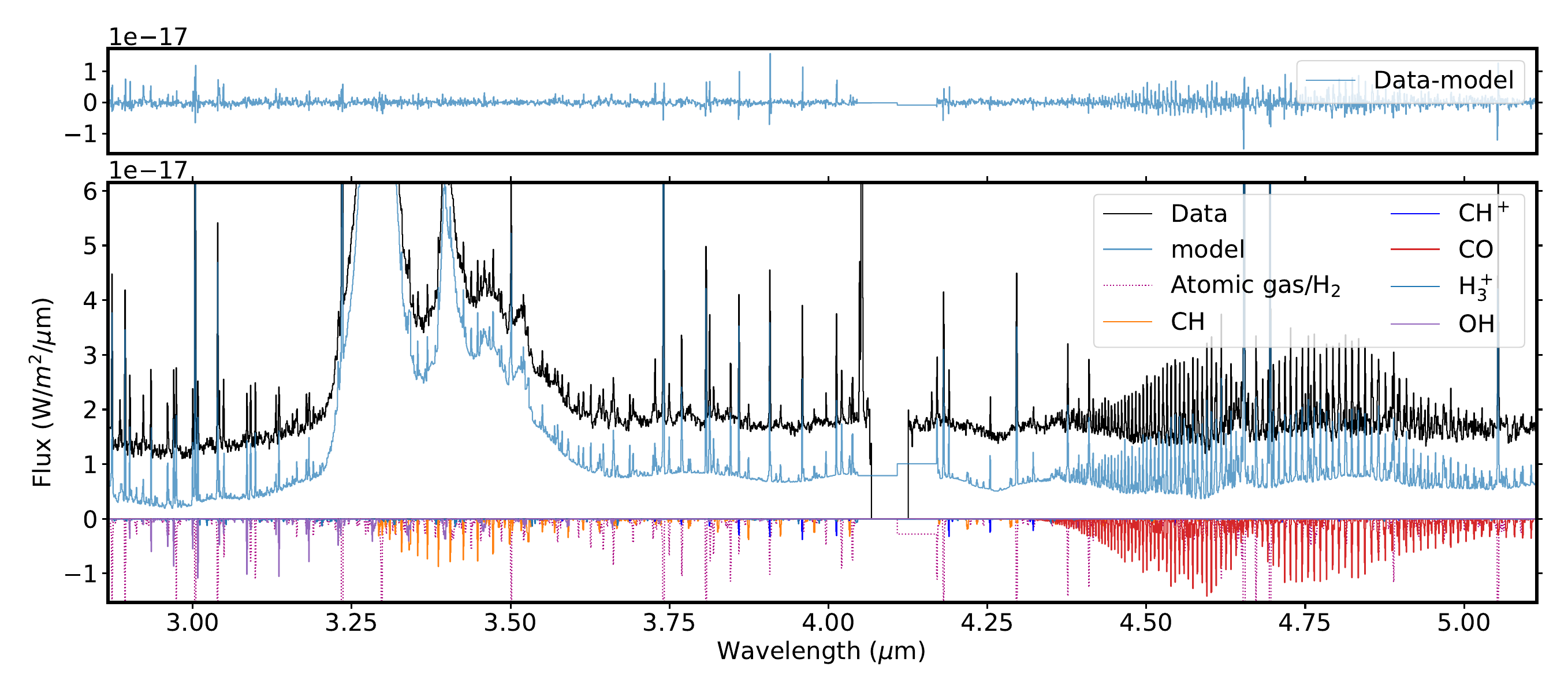}
\caption{\textit{Bottom panel: }\textsc{PAHTATmol} shifted fit result (blue curve) on the NIRSpec f290lp filter spectrum of d203-506 (black curve). The different colors at the bottom of the panel show the inverted individual molecular spectra, and a dotted {magenta} line shows the atomic and H$_2$ emission lines.
\textit{{Upper} panel: }Model residual (observed spectrum minus the model).}
\label{fig:d203-506_ns}
\end{figure*}

\section{JWST spectroscopic observations of d203-506}
\label{section:observations}
 d203-506 is 
situated in the line of sight toward the Orion Bar \citep{bally2000disks}, at about 0.25 pc from the massive Trapezium stars that are at a distance of $\sim390$~pc \citep{apellaniz2022villafranca}, inside the Orion Nebula. 
\mbox{d203-506} is about 100 au in radius, and has an estimated mass {of} $\sim 10~M_{ \rm Jup}$ as well as a central star estimated mass of  $M_{\star}=0.2 \pm 0.1 M_{\odot}$ \citep{BerneO_2024_science}. The central star is obscured by this nearly edge-on disk \citep{bally2000disks}. 
\mbox{d203-506} was observed as part of the JWST ERS program PDRs4All \citep{PDRs4All} with
NIRSpec-IFU \citep{boker_nirspec_ifu_2022}. The data reduction is presented in \citet{peeters2024pdrs4all}. 
In order to obtain a  high-quality spectrum of \mbox{d203-506}, we extracted a spectrum from the NIRSpec cube using an aperture of 0.1\arcsec\ (corresponding to 39 au at 390~pc) centered on the position of the bright spot ($RA=$5$^h$35$'$20.314$''$, $DEC=$-5$^\circ$25$'$05.47$''$) on the northwestern tip of the disk facing the illuminating star, $\theta^1$ Ori C.  
The resulting spectrum is presented in Figs.~\ref{fig:d203-506_ns} and \ref{fig:d203-506_ns_all}. 
As we are interested in simple {molecular} ions and radicals in this study, we only focus on the f290 filter, spanning from 2.87 to 5.2~\um, which is where many fundamental stretching 
bands are present.  

\section{Fitting of the spectrum}
\label{section:fitting_main}

\subsection{Fitting procedure} 

In Appendix~\ref{section:pahtatmol} we present the \textsc{PAHTATmol} tool, which allows  a global molecular model to be fit to an observed JWST spectrum of an irradiated disk. Briefly, this model includes broad individual Gaussians to reproduce the emission from polycyclic aromatic hydrocarbons (PAHs), blackbodies to reproduce the stellar and disk dust continuum emission, and the ro-vibrational emission from multiple molecular species. 
Here we include emission from 19 species, which are listed in Table~\ref{table:molecule_content}. 
This list includes simple species previously observed in disks such as CO, OH, \ch{HCO+}, and \ch{CH+}, but also a number of other species that have not yet been detected in disks (e.g.,\ \ch{CH,~H3+}), and species not detected in space at all (e.g.,\ \ch{~C2H2+}).
For all species we used state-of-the-art {public} spectroscopic data and computed their global emission assuming local thermodynamical equilibrium (LTE), {meaning that the ro-vibrational levels follow a Boltzmann distribution at a single excitation temperature ($T_{\rm ex}$).\footnote{{In many cases, however, the infrared emission arises from \mbox{non-LTE} excitation {($T_{\rm ex}$\,$\neq$\,$T_{\rm gas}$)}—for instance, driven by radiative or chemical formation pumping—which requires more sophisticated models that explicitly solve the statistical equilibrium equations, incorporating the IR-to-UV radiation field and both chemical formation and destruction  processes (e.g., \citealt{Zannese_24_oh,zannese2025} for  \ch{OH}, \ch{CH3+}, and \ch{CH+}).}}}
\mbox{\textsc{PAHTATmol}} provides a first-order simulation, allowing a simultaneous fit over the full JWST wavelength range that includes all potential ro-vibrational lines (blended or not) of the species that might be present.
Thus, \mbox{\textsc{PAHTATmol}} provides LTE column densities ($N)$ and excitation temperatures 
$T_{\rm ex}$, assuming \mbox{single-slab} radiative transfer. 
In protoplanetary disks, gas densities are typically high enough that, {for many molecular species, the rotational temperature ($T_{\rm rot}$) of the \mbox{low-lying} rotational levels in the ground vibrational state tends to approximate the gas temperature,
with $T_{\rm rot}$\,$\lesssim$\,$T_{\rm gas}$.}\,
For instance, the infrared \ch{H2} \mbox{rotational} emission observed in the irradiated layers of \mbox{d203-506}, with {$T_{\rm rot}$\,$\simeq$\,1000\,K}\, \citep{BerneO_2024_science}, provides a good measure of the aperture-averaged gas temperature.

\subsection{Fit results} 

Figure~\ref{fig:d203-506_ns} shows the resulting best-fit model using \textsc{PAHTATmol} over the 2.87 to 5.2 \um\ wavelength range (corresponding to NIRSpec f290lp filter). In this range, we are able to reproduce the vast majority ($\geq90\%$) of the lines present in the data.\footnote{ 
The spikes in the residual (difference) spectrum (see the upper panel of Fig.~\ref{fig:d203-506_ns}) arise from strong emission lines, caused by inaccuracies in the simplified modeling of the excitation conditions.}
In  particular, we report the detections of \ch{CO,~CH,~CH+ ,~OH, and~H3+}.
 We note that the estimated excitation temperatures, hereafter $T$, provided by the \textsc{PAHTATmol} fit are to be taken as  indicators rather than precise values, and are given with an error of $\sim \pm 300$~K. For both CO and \ch{H3+}, the best-fit temperatures are a combination of $T= 1000$~K and 2000~K. This implies that two components are needed to fit the data, likely tracing the molecule's presence along a temperature gradient. 
 For all species, 
 the derived temperatures are  high ($\geq1000$~K), indicating their presence in 
 \mbox{UV-irradiated} hot gas {and-or excitation by  formation or radiative pumping}\, \citep[e.g.,][]{zannese2025}.

\section{New detections in a protoplanetary disk}
\label{section:ch_main}
\begin{figure}[t]
\centering
\includegraphics[width=0.45\textwidth]{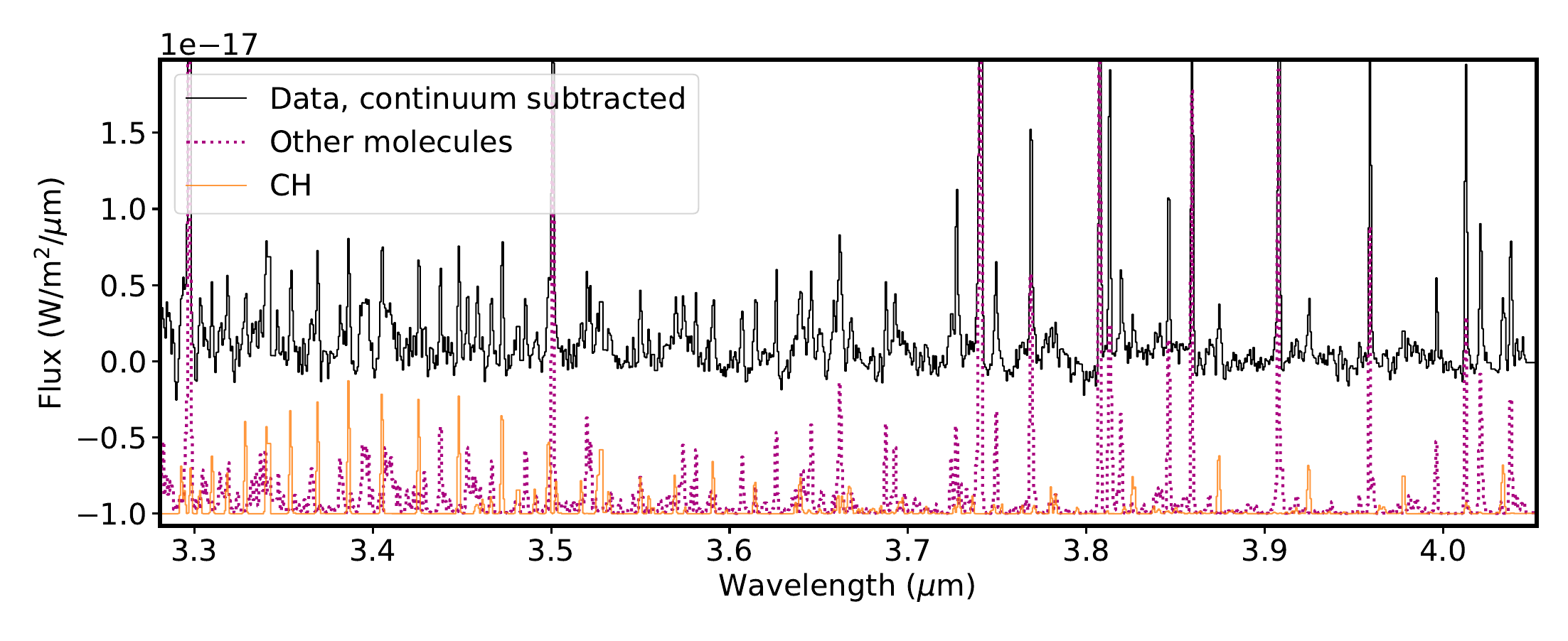}
\caption{Near-infrared \ch{CH} rovibrational emission in d203-506. {The continuum-subtracted observations are shown in black, and the CH model—offset for clarity—is shown in orange}.
The dotted {magenta} lines indicate model spectra for additional species.}
\label{fig:ch}
\end{figure}

In addition to the previously detected species \ch{CO,~OH,\,and~CH+}, we used \textsc{PAHTATmol} to report, {according to our detection threshold definition (see Appendix~\ref{section:fitting})}, the first detections of \ch{CH} (see Fig.~\ref{fig:ch}) and  \ch{H3+} (see Figs.~\ref{fig:h3p} and \ref{fig:h3p_branches}) in a protoplanetary disk. In the following we discuss the implications of these detections.

\subsection{Methylidyne radical: CH}

The methylidyne radical has been detected in the interstellar medium and has been used as a powerful tracer of molecular hydrogen {and} of the initial steps of hydrocarbon chemistry \citep[e.g.,][]{gerin2010interstellar}. We detect the CH ro-vibrational emission lines around 3.3~\um in the same wavelength range of the \mbox{3.3-3.4~\um} PAH emission bands (C--H stretching).
The detected CH emission lines correspond to both the R branch ($<3.62$~\um) and the P branch  ($>3.68$~\um). 
For CH, we find a  temperature 
{$T\simeq 2000$~K and a column density of 4.5$\times 10^{14}$~\cmsq}.
Using the H$_2$ column density derived by \citet{BerneO_2024_science} and assuming $N$(H)\,=\,$N$(H$_2)$ in the disk PDR, we infer a CH abundance of a few 10$^{-7}$ with respect to H nuclei.

{The bottom left panel of Fig.~\ref{fig:ch_map} shows the spatial distribution of the CH 3.426\,$\mu$m emission line. This emission is slightly less extended than that of vibrationally excited H$_2$ (e.g., as traced
by the $v=1-0$ $S$(1) line emission) 
and implies that CH arises from deeper layers of the disk PDR. 
Our photochemical models (see Appendix~\ref{Appendix-PDR-model}) show that, as for CH$^+$ and CH$_3^+$, CH is a natural product of simple hydrocarbon photochemistry
(see the chemical sketch in \mbox{Fig.~11} of \citealt{goicoechea2025pdrs4all})
triggered by the {presence of C$^+$ ions, C atoms}, elevated temperatures, and the enhanced reactivity of  excited H$_2$ \citep{BerneO_2024_science}. 
{Compared to \ch{CH+} and \ch{CH3+}, the model suggests that most of the CH column density originates from slightly deeper layers within the PDR (see Fig.~\ref{fig:CH_models}), where CH is formed through the dissociative recombination of \ch{CH3+} and through reactions of atomic carbon with vibrationally excited \ch{H2}.}
\mbox{Figure~1} of \citet{Berne_ch3p} shows that the IR CH$^+$ and CH$_3^+$ emission peak toward the outermost irradiated disks layers (the bright spot), whereas CH peaks deeper inside the disk PDR (Fig.~\ref{fig:ch_map}).
Our {photochemical} model predicts $N$(CH)\,$\lesssim$\,10$^{14}$\,cm$^{-2}$, which is {a factor of $\lesssim$\,5 lower 
than the estimated LTE column density. }
{This might suggest that other chemical routes contribute to CH formation—such as its indirect production through PAH or carbonaceous grain photodestruction.
Alternatively, reactions of electronically  excited far-UV \mbox{(FUV)-pumped} C atoms with H$_2$ may play a role \citep[e.g.,][]{Lezana07,Goicoechea2024}, or, more likely, radiative pumping contributes to CH excitation \citep[see Appendix of][for \ch{CH3+}]{Berne_ch3p} as CH exhibits strong electronic transitions in the visible and FUV \citep[e.g.,][]{Oka13}, which can be excited by external starlight.}

\begin{figure}[t]
\centering
\includegraphics[width=0.48\textwidth]{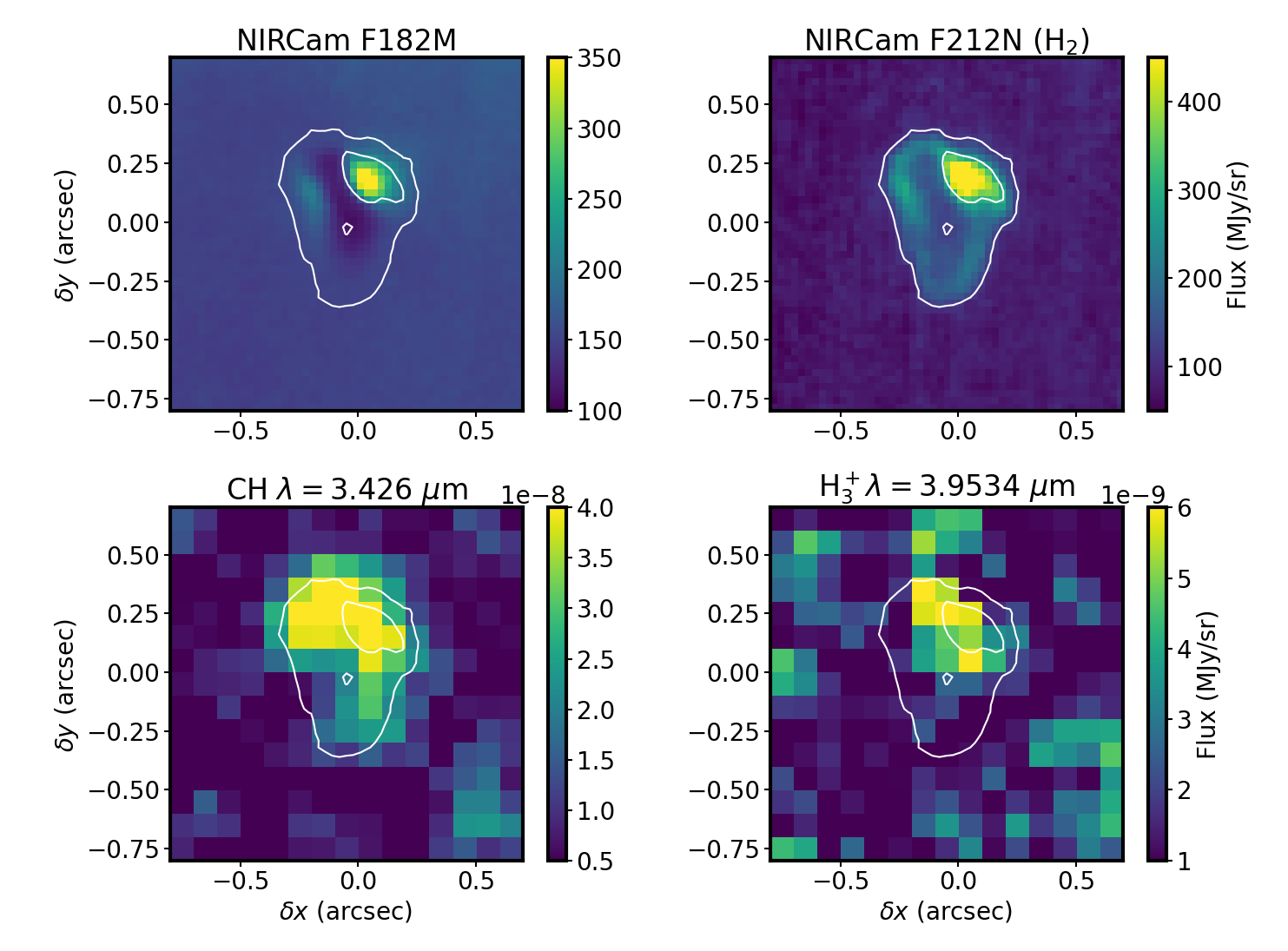}
\caption{IR JWST images of d203-506. {\it Top left: }NIRCam F182M (broad filter centered at 1.82~\um). {\it Top right: }NIRCam 
F212N (\ch{H2} 1-0 $S$(1) emission). 
{\it Bottom left: }NIRSpec \ch{CH} emission map, continuum subtracted. {\it Bottom right:} NIRSpec $Q$(1,0) \ch{H3+} emission map, continuum subtracted. All panels show F212N contours in white.}
\label{fig:ch_map}
\end{figure}

\subsection{Trihydrogen cation: \ch{H3+}}
   
The trihydrogen cation, \ch{H3+}, is the electronically simplest stable polyatomic molecule. {Figure~\ref{fig:h3p} shows the {likely} detection of several H$_{3}^{+}$ emission lines from the \( \nu_{2}=1 \rightarrow 0 \) band, including the $R$(1,0), $Q$(1,0), {$Q$(5,$G$)\footnote{$G$=0, 1, 2, and 3}}, 
$R$(7,0)$^l$, and $R$(7,6)$^u$ lines. 
{The reported lines are weak and lie near the detection threshold, 
{but the stacked spectrum clearly shows an emission feature
(see details in Appendix~\ref{section:h3p})}.}
\mbox{Table~\ref{table:h3p_molecule}} summarizes the transition properties.
Other expected \ch{H3+} lines are blended with emission lines from other abundant species, 
{leaving the presence of  \ch{H3+} emission in \mbox{d203-506} open to interpretation}.
In Fig.~\ref{fig:h3p_branches}, we show the complete best-fit
\ch{H3+} emission model for each \( \nu_{2}=1 \rightarrow 0 \) branch 
($R$, $Q$, and $P$) and compare it with the observed {minus source-model} spectrum.
\citet{pereira2024h3+} recently reported the   detection of \ch{H3+}  (in emission and absorption) in the interstellar medium of galaxies using JWST-NIRSpec.
The higher-energy \ch{H3+} lines {reported} in \mbox{d203-506} (\mbox{Fig.~\ref{fig:h3p}}) are not observed in the galaxies, suggesting that the excitation conditions in the disk—specifically the gas temperature and density—are higher. Interestingly, some of these excited lines, for instance
the {$Q$(5,0)} line,
{are} detected in the ionosphere of Jupiter
\citep[][]{Oka90}.
Our detection {would} represent the first {identification} of H$_{3}^{+}$ in a protoplanetary disk. 

The H$_{3}^{+}$ emission in \mbox{d203-506} peaks near the bright spot (see bottom right panel of Fig.~\ref{fig:ch_map}), which also coincides with the peak of near-IR H$_2$ $v$\,=\,1-0 $S$(1) emission. This region corresponds to gas that is strongly illuminated by FUV radiation. The H$_{3}^{+}$ emission also overlaps with the emission peak of highly excited rotational lines of OH, a clear indication of water vapor photodissociation \citep{Zannese_24_oh}. 
We find  \mbox{$T\geq 1000$~K} and a column density of 1.1$\times 10^{13}$~\cmsq, thus an abundance
of a few 10$^{-8}$, using the same circular surface area as for CH.
{Given the very large radiative transition probabilities of the \ch{H3+} ro-vibrational emission lines (on the order of $\sim$100\,s$^{-1}$), these lines cannot be thermalized ($T_{\rm vib}$\,$\ne$$T_{\rm gas}$) by inelastic collisions   at the densities of $\sim$10$^{7}$\,cm$^{-3}$ expected in the disk PDR \citep[e.g.,][]{BerneO_2024_science}. 
It is therefore important to distinguish between the best-fitting LTE temperature, which approximately reflects the rotational excitation, and the vibrational excitation temperature.
{{The presence} of vibrationally excited H$_3^+$ emission in {protoplanetary disks} {raises important}  questions regarding  the mechanisms driving its formation and the excitation of these levels—whether through chemical pumping, radiative pumping, or both.}
In a companion paper, \citet{Goicoechea2025_h3p} find that H$_3^+$ can efficiently form in strongly irradiated ($G_0$\,$>$\,10$^3$) dense gas through chemical routes 
{involving a hot water vapor reservoir,} 
triggered by UV radiation, 
and largely independent of the cosmic-ray ionization rate. For d203-506, they predict a H$_3^+$ column density of $\lesssim$10$^{13}$\,cm$^{-2}$ in the disk PDR, in good agreement with the value estimated here.
}

\begin{figure}[h!]
\centering
\includegraphics[width=0.42\textwidth]{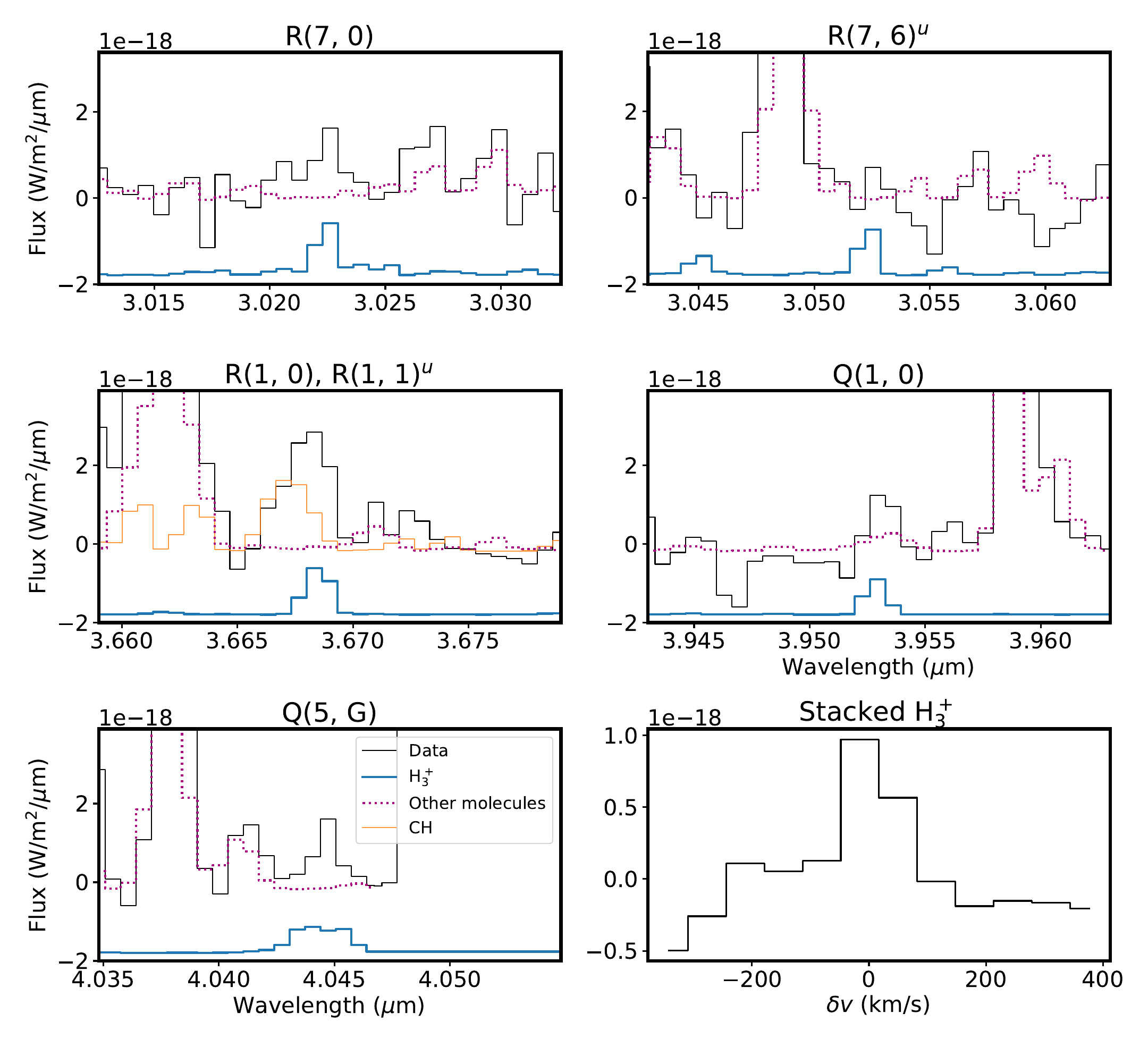}
\caption{{\ch{H3+} lines  in d203-506. The label of each (centered) transition is indicated in the title of each panel, in the format Branch($J, G$). In the bottom left panel, transitions with $G = 0$, 1, 2, and 3 are superimposed. Continuum-subtracted observational data are shown in black, centered on each \ch{H3+} line. LTE models for \ch{H3+} (blue), \ch{CH} (orange), and other species (dotted {magenta}) are overplotted. \ch{H3+} model  has been vertically offset for clarity (see also Fig.~\ref{fig:h3p_branches})}.
{The lower right panel shows the stacked \ch{H3+} spectrum in velocity space
(see Appendix~\ref{section:h3p})}.
}
\label{fig:h3p}
\end{figure}

\section{Conclusion {and outlook}}
\label{section:conclusion}

{{{We claim the detection of} two new molecules in a protoplanetary disk: the methylidyne radical (CH) and the trihydrogen cation (\ch{H3+})}. The detection of CH, which shows a slightly different spatial emission distribution compared to that of CH$^+$ and CH$_3^+$ \citep{Berne_ch3p}, is consistent with the expectations of PDR hydrocarbon chemistry driven by the high gas temperature and enhanced reactivity of excited H$_2$, which efficiently converts C$^+$ ions and C atoms into simple hydrocarbons.} 
{The presence of infrared H$_3^+$ emission in the disk's PDR component, {if confirmed, would} suggest that chemical formation routes unrelated to cosmic-ray ionization contribute to its formation.

{The detection of hot \mbox{vibrationally} excited \ch{CH} and \ch{H3+} 
underscores the need to explore the \mbox{non-LTE} excitation mechanisms of these species,
{which will be the subject of future papers}.}\,
 All in all, we expect these two new molecules to be present in the IR spectra of other strongly irradiated protoplanetary disks.
\begin{acknowledgements}
We thank the referees for her/his comments and suggestions, which helped improve the clarity of the manuscript.
This work is based [in part] on observations made with the NASA/ESA/CSA \textit{James Webb} Space Telescope. The data were obtained from the Mikulski Archive for Space Telescopes at the Space Telescope Science Institute, which is operated by the Association of Universities for Research in Astronomy, Inc., under NASA contract NAS 5-03127 for JWST. These observations are associated with programs \#1288.
IS, OB are funded by the Centre National d'Etudes Spatiales (CNES) through the APR program.
This research received funding from the program ANR-22-EXOR-0001 Origins of the Institut National des Sciences de l’Univers, CNRS. 
This project is co-funded by the European Union (ERC, SUL4LIFE, grant agreement No101096293). AF also thanks project PID2022-137980NB-I00 funded by the Spanish Ministry of Science and Innovation/State Agency of Research MCIN/AEI/10.13039/501100011033 and by “ERDF A way of making Europe”.
C.B. acknowledges support from the Internal Scientist Funding Model (ISFM) Laboratory Astrophysics Directed Work Package at NASA Ames (22-A22ISFM-0009) and is grateful for an appointment at NASA Ames Research Center through the San Jos\'e State University Research Foundation (80NSSC22M0107).
\end{acknowledgements}}

\appendix

\section{\textsc{PAHTATmol}}
While atomic emission lines are quite straightforward to detect in astronomical spectra due to their temperature independent line ratios and well known line positions, the detection of molecules requires more intensive modeling, since their relative line intensities and thus emission pattern depends on the excitation temperature. A complete spectroscopic model, derived most often from laboratory spectroscopic studies, is needed to predict the emission spectra at different temperatures. Once this is known, specific molecules can be identified in an astrophysical object, and their excitation temperature and abundance can be derived from fitting the relative and absolute intensities to the observations.
With the new JWST spectroscopic data allowing for fine molecular detection and even the discovery of new molecules like \ch{CH3+} \citep{Berne_ch3p, ChangalaB_ch3p}, being able to quickly find the presence of every available molecule in the observation is of great importance.
We aim to provide a tool to search for the different molecules across JWST's wavelength coverage ($0.9 - 28$~\um) and also derive the column densities of those detected species, if the geometry of the system is known\footnote{The emission size and the distance of the object.}.
This tool is called \textsc{PAHTATmol} and focuses on small molecules, radicals and ions in the UV-irradiated environment but also includes atoms, neutral and ionized.

\label{section:pahtatmol}
\textsc{PAHTATmol} stands for Polycyclic Aromatic Hydrocarbons Toulouse Astronomical Templates with MOLecules. 
This tool is based on the \textsc{PAHTAT} tool \citep{pilleri_evaporating_2012, Foscino_19}, which fits spectra using linear methods. 
It assumes that each PDR spectrum is a linear combination of continuum, emission lines, AIBs (Aromatic Infrared Bands) and noise. 
In \textsc{PAHTATmol}, we include the contribution from molecules.

\subsection{Computation of molecular emission spectra}

In \textsc{PAHTATmol}, we focus on small molecules, radicals and ions in the UV-irradiated environments. 
Molecules are chosen using the most probable chemical paths in such environment. 
The 19 molecules focusing on hydrides, ions and radicals and are the following: 
\ch{CH}, \ch{CH+}, \ch{OH}, \ch{OH+}, \ch{NH}, \ch{HCl}, \ch{CO}, $^{13}$\ch{CO}, \ch{H3+}, \ch{H3O+} \ch{SH}, \ch{HeH+}, \ch{HCO+}, 
\ch{SiH}, \ch{HCN+}, \ch{HCNH+}, \ch{C3H+}, \ch{C2H2+}, and \ch{CH3+}. 
To obtain an estimation of the excitation temperature of each species, we need to model their emission at different temperatures. In our case, the temperature range is from 300 to 2000~K.
To fit the spectrum using a linear combination of molecular emissions we generate, for each molecule, one spectrum at each temperature over the wavelength range of 0.9 to 27~\um\ to match JWST NIRSpec and MIRI MRS.
The \textsc{pgopher} program is used for this. 

\begin{table}[h!]
\centering
\caption{Molecules included in PAHTATmol, with references to the spectroscopic 
parameters used to compute the emission spectra. }
\label{table:molecule_content}
\begin{tabular}{lcc}
\hline
Species           &    Band             & source  \\
\hline                                              
\ch{CH}           &    $\nu_1$          &  7,15  \\
\ch{CH+}          &    $\nu_1$          &  5  \\
\ch{OH}           &    $\nu_1$            &  7,10,11  \\
\ch{OH+}          &    $\nu_1$          &  6,7  \\
\ch{NH}           &    $\nu_1$            &  22  \\
\ch{HCl}          &    $\nu_1$            &  16  \\
\ch{CO}           &    $\nu_1$            &  12,13  \\
$^{13}$\ch{CO}    &    $\nu_1$            &  12  \\
\ch{H3+}          &    $\nu_2$ &  8,9  \\
\ch{H3O+}         &    $\nu_1$, $\nu_2$, $\nu_3$, $\nu_4$  &  1,2,3  \\
\ch{SH}           &    $\nu_1$          &  17  \\
\ch{HeH+}         &    $\nu_1$            &  14  \\
\ch{HCO+}         &    $\nu_1$, $\nu_2$, $\nu_3$   &  4, 18  \\
\ch{SiH}         &    $\nu_1$            &  21  \\
\ch{HCN+}          &    $\nu_1$          &  to be submitted  \\
\ch{HCNH+}        &    $\nu_1$, $\nu_2$ &  19,20  \\
\ch{C3H+}         &    $\nu_1$,$\nu_2$            &  25,26  \\
\ch{C2H2+}        &    $\nu_3$, $\nu_5$            &  27,28,29  \\
\ch{CH3+}         &    $\nu_2$, $\nu_3$, $\nu_4$ &  23,24  \\
\hline
\end{tabular} \\
\tablefoot{The modes included in the model are 
also indicated. 
1: \citet{gruebele1987study}; 2: \citet{liu1985infrared}; 3: \citet{tang1999infrared};  
4: \citet{lattanzi2007rotational}, 
5: \citet{Domeneche_18_CHp};
6: \citet{hodges2017fourier}; 7: \citet{bernath2020mollist}; 
8: \citet{mizus2017exomol}; 9: \citet{bowesman2023exomol}; 10: \citet{brooke2016line}, 11: \citet{yousefi2018new};
12: \citet{li2015rovibrational}; 13: \citet{somogyi2021calculation}; 14: \citet{Crofton_89_hehp};
15: \citet{masseron2014ch}; 16: \citet{gordon2017hitran2016}; 17: \citet{gorman2019exomol}; 
18: \citet{Siller_13_hcop};
19: \citet{altman1984high}; 20: \citet{altman1984observation};
21: \citet{yurchenko2018exomol}; 22: \citet{perri2024full}; 23: \citet{Berne_ch3p}; 24: \citet{ChangalaB_ch3p};
25: \citet{bast2023ro}; 
26: \citet{schmid2022leak};
27: \citet{schlemmer2024high};
28: \citet{jagod1992infrared}; 
29: \citet{asvany2005experimental}.
}
\end{table}

\textsc{pgopher} \citep{western2017:PGOPHER} is a versatile program designed for the simulation and fitting of rotational, vibrational, and electronic spectra of molecules. 
It allows to generate a LTE spectrum of a molecule from either molecular constants or from provided state and transition files.
The aim of using \textsc{pgopher} is to generate, for each molecule cited previously, five spectra corresponding to temperatures of 300, 500, 1000, 1500 and 2000~K, respectively.
Each spectrum is generated using either states and transition files from the \textsc{exomol} database \citep{Exomol_ref}  or from molecular constants from the literature.
In \textsc{pgopher}, each spectrum is generated between 350-12000~cm$^{-1}$ (0.9-28~\um) and convolved by a Gaussian with a FWHM of 0.3~cm$^{-1}$, corresponding to NIRSpec resolution, and sampled with at least 100 000 points. 
In addition, we set the spectra intensity units to W$/$molecule, which will allow a column density estimate for each detected molecule.
The generated spectrum is then exported and split into two parts: NIRSpec and MIRI/MRS wavelength ranges, resampled to each instrument's spectral resolution. 
Finally, we add each spectrum into the \textsc{PAHTATmol} library.  
The list of molecules and references for the adopted spectroscopic data are given in Table~\ref{table:molecule_content}.

\subsection{Fitting a spectrum with \textsc{PAHTATmol}}
\label{section:fitting}
Given a JWST spectrum, we can now fit the spectrum the same way as \textsc{pahtat} but adding the generated molecule spectra. 
As in \citet{Foscino_19} we use a non negative least square algorithm to estimate each molecule contribution. 
We choose to use a linear combination of molecule contribution because it is faster than non linear methods. 
The resulting fit gives us factors needed to reproduce the data for each molecule.
From these factors and the residue of the fit (data-model), we can assess if one molecule is detected based on multiple parameters.

To assess if a molecule is detected, we first select the molecules with a non null coefficient from the results.
We then compute the ratio between the molecule emission positions times the fitted factor $F_{\rm mol}$ and divide by the standard deviation of the residuals at the same position across 20 spectral pixels. 
This gives us a signal to noise ratio (S/N). This S/N formula is given by
\begin{equation}
    {\rm S/N} = \sqrt{\sum_n \left ({F_{\rm mol} \times I_{{\rm peak}_n} \over{\sigma_{{\rm peak}_n} [\pm10]}}\right )^2},
\end{equation}
with $\rm peak_n$ being the $n$'s peak position and $I_{\rm peak_n}$ the intensity of the $n$'s peak, and $\sigma_{{\rm peak}_n} [\pm10]$ the standard deviation of the spectrum to fit ranging ten spectral pixels and centered at the peak position.
We define the detection threshold as \mbox{S/N > 5}. 
{According to our definition},
we clearly detect \ch{CH}, \ch{CH+}, \ch{CO}, \ch{H3+} and \ch{OH}. For CH we obtained S/N = 48, for \ch{CH+}  S/N = 13, for CO S/N of 40, and for both \ch{H3+} and OH S/N = 8.
For each detected molecules, $F_{\rm mol}$ provides an estimation of its {LTE} column density with \mbox{$N\approx F_{\rm mol} \times d^2 / S$}, taking into account the surface $S$ of the extracted spectrum aperture and the distance $d$ to the object.

\begin{figure*}[h!]
\centering
\includegraphics[width=0.9\textwidth]{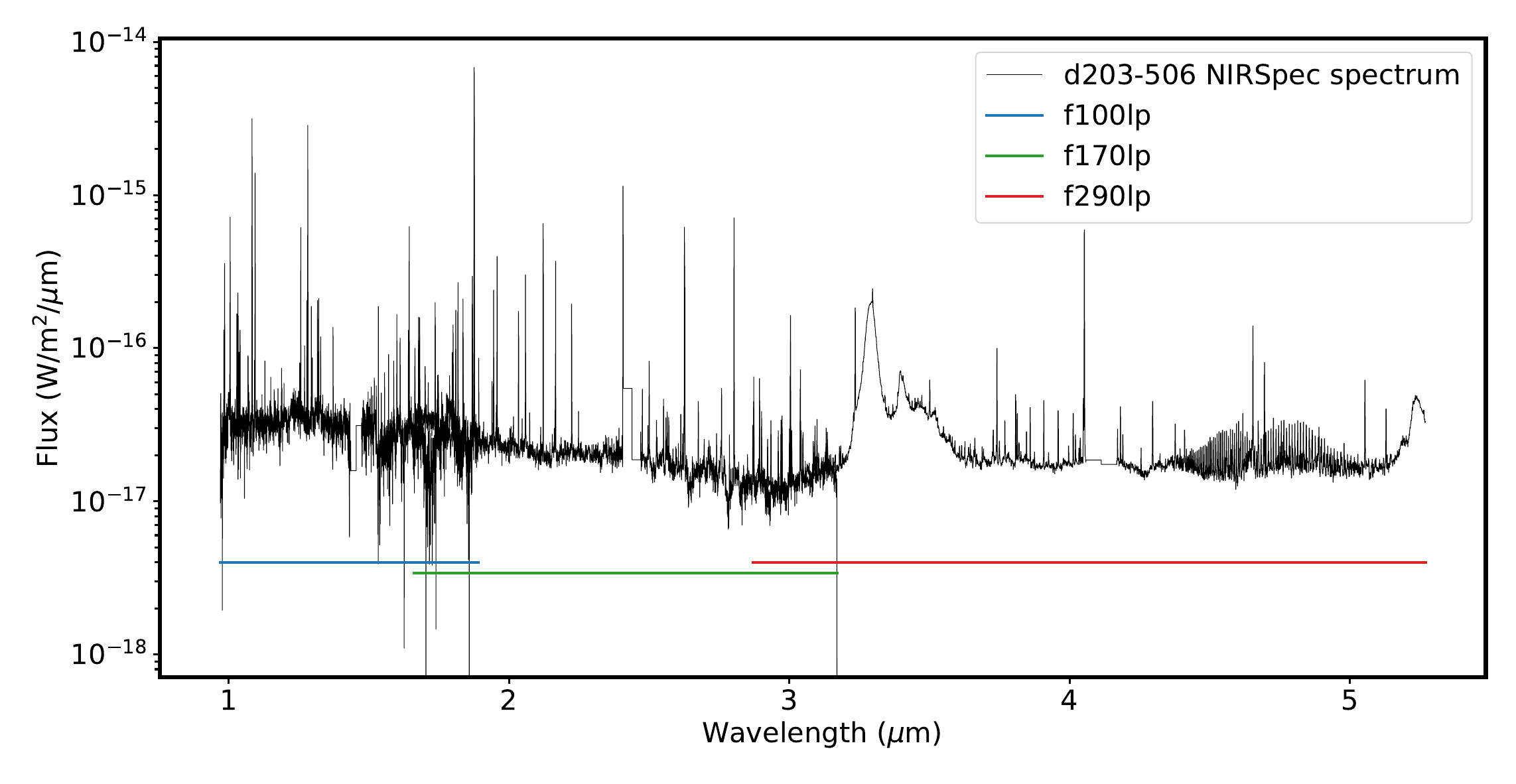}
\caption{d203-506 NIRSpec spectrum in black. Each NIRSpec filter wavelength coverage is shown below the spectrum.}
\label{fig:d203-506_ns_all}
\end{figure*}

\section{Spectroscopy of H$_3^+$}
\label{section:h3p}

H$_{3}^{+}$, with two electrons, is the electronically simplest stable polyatomic molecule. 
Having the symmetry of an equilateral triangle with no permanent dipole moment, 
it does not exhibit a rotational spectrum. 
Also, H$_{3}^{+}$ is unstable in its singlet electronically excited state, 
and no sharp electronic spectrum is known \citep[e.g.,][]{Oka13}.
{However, it} has two vibrational modes $\nu_1$ and $\nu_2$, 
of which the latter is IR active. 
It was first detected in the laboratory by \cite{Oka80},
with a band origin at $\sim$4\,\um\ (2521.6~cm$^{-1}$).

The rotational states of the ground vibrational state are characterized by two quantum numbers: the total angular momentum \( J \) and its projection onto the symmetry axis \( K \). Since the proton is a fermion, the Pauli exclusion principle requires that all states be antisymmetric under the exchange of hydrogen nuclei. As a result, the rotational ground state \( (J = 0) \) is forbidden, making \( J = 1 \) the lowest possible state for H$_3^+$.  H$_3^+$ exists in two spin isomers, distinguished by the total nuclear spin \( I \): \textit{ortho}-H$_3^+$ with \( I = 3/2 \) and the quantum number \( K = 3n \), and \textit{para}-H$_3^+$ with \( I = 1/2 \) and \( K = 3n \pm 1 \).
For the \( \nu_2 \) excited state, an extra quantum number is needed for the vibrational angular momentum, \( l = \pm 1 \), in addition to \( J \) and \( K \). While the parity is given by \( (-1)^K \), as in the ground state, the ortho and para labels depend on whether \( K - l \) is a multiple of 3. Therefore, a quantum number \( G \equiv |K - l| \) is used for convenience (for details and spectroscopic constants see \citealt{mizus2017exomol}). For the ground state, we have \( l = 0 \), \( G = K \).
The selection rules for the \( \nu_{2} = 0\rightarrow 1 \) ro-vibrational transitions are \( \Delta J = 0 \) or \( \pm 1 \), and \( \Delta G = 0 \) 
\citep[e.g.,][]{Oka13,Miller20}. In this work we detect the following lines: \mbox{R(7, 0)}, \mbox{R(7, 6)$^u$}, \mbox{R(1, 0)}, \mbox{Q(1, 0)} and {\mbox{Q(5, G)$^l$} (G=0, 1, 2 and 3)} (notations are from \citealt{lindsay2001comprehensive}) from the $\nu_2$ band (see Fig~\ref{fig:h3p}). 
Those lines are the ones that are not blended with other emission lines (apart from \mbox{R(1, 0)} for which a CH emission line is {partially} blended with it). 
{To increase the line sensitivity and support the detection of \ch{H3+}, we looked for a line emission feature in the stacked spectra. 
We selected wavelength windows around the \ch{H3+} lines listed in \mbox{Table~\ref{table:h3p_molecule}}, as well as  the $R(3,3)^l$ and $Q(3,0)$ lines, centered at 3.534 and 3.986~\um, respectively. All these lines are expected to have similar intensities.
We resampled them to the same channel resolution and then stacked them.
Owing to line crowding from multiple molecules in the  NIRSpec spectrum of \mbox{d203-506}, we stacked the residual spectrum (observed minus PAHTATmol model), excluding \ch{H3+}. The lower-right panel of \mbox{Fig.~\ref{fig:h3p}} shows the resulting  spectrum, which clearly displays an emission feature at the expected position of  \ch{H3+} lines (zero velocity).}

We recall that our global PAHTATmol model fit of \mbox{d203-506's} spectrum includes all ro-vibrational lines (blended or not) of all considered species.

\begin{figure*}[h!]
\centering
\includegraphics[width=0.9\textwidth]{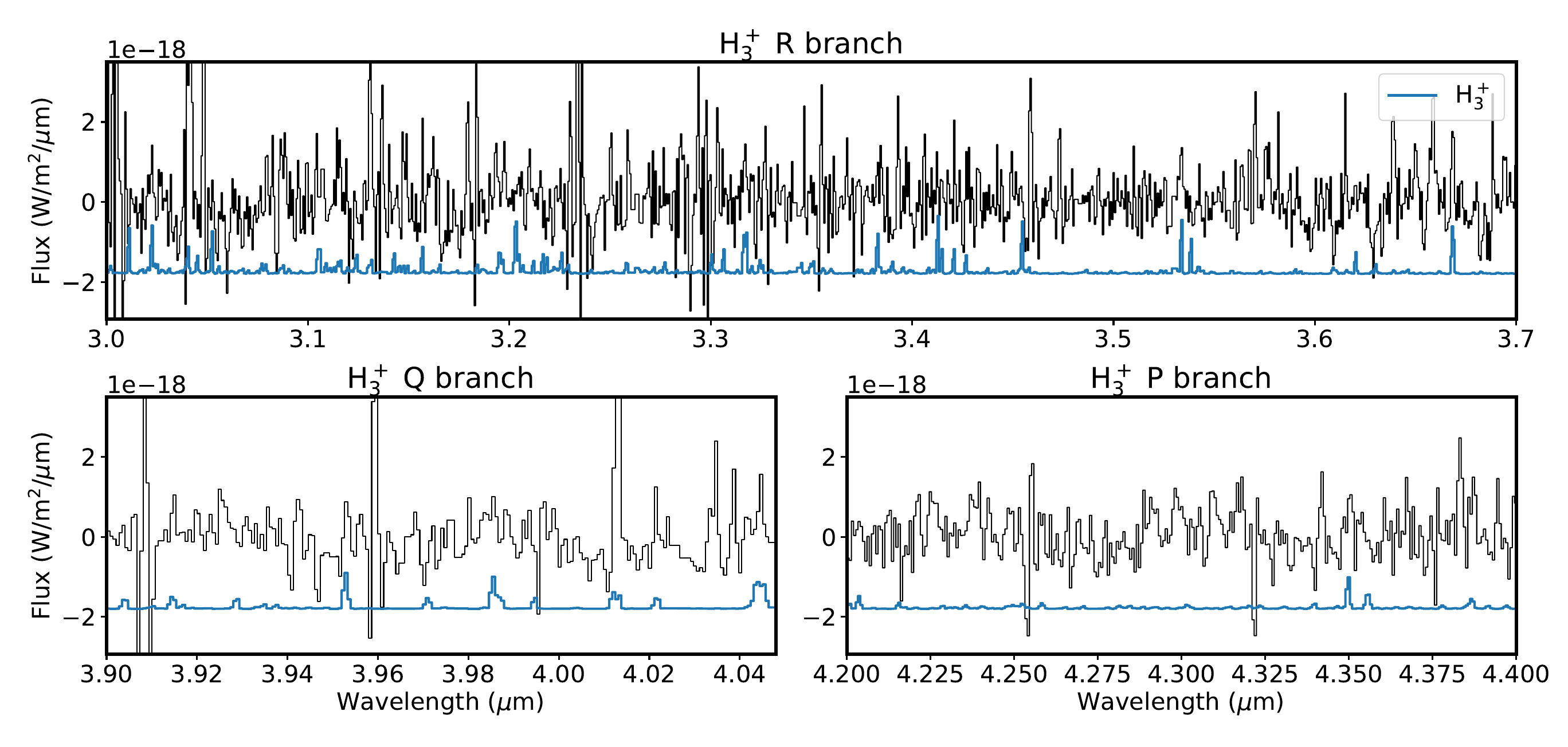}
\caption{\ch{H3+} R--, Q--, and P--branch wavelength ranges from top to bottom. 
In each panel, the data are shown in black and have been model-subtracted (including continuum, AIBs, atoms, and molecules, excluding \ch{H3+}), while the \ch{H3+} model is overplotted in blue and vertically offset. }
\label{fig:h3p_branches}
\end{figure*}

\begin{table*}[h!]
\centering
\caption{\ch{H3+} emission line properties.}
\label{table:h3p_molecule}
\begin{tabular}{lclcccc}
\hline
Wavenumber & wavelength & Transition  & $E_{\rm low}$ &  $E_{\rm up}$  &   A & $F_{\rm Obs}$ \\
 (cm$^{-1}$)& (\um) &    &  (K) &   (K)  &   (s$^{-1}$) & W m$^{-2}$ \\
\hline
3308.685 (05)& 3.022 & $R$(7,0)      &  3338.5  &   8098.9  &     174.13 & 7.2$\times 10^{-22}$ \\
3276.197 (05)& 3.052 & $R$(7,6)$^u$  &  2282.7  &   6996.5  &     100.62 & 3.8$\times 10^{-22}$ \\
2726.220 (05)& 3.668 &   $R$(1,1)$^u$&   92.2   &   4014.7  &       59.7 & Blended with $R$(1,0)\\
2725.898 (05)& 3.668 & $R$(1,0)      &   125.1  &   4047.07 &      97.91 & 2.4$\times 10^{-22}$ \\
2529.724 (05)& 3.953 & $Q$(1,0)      &   125.1  &   3764.8  &     127.75 & 2.3$\times 10^{-21}$ \\
2472.325 (10)& 4.044 & $Q$(5,1)$^l$  &  1798.9  &   5356.0  &     108.56 & 1.7$\times 10^{-21}$ \\
\hline
\end{tabular}\\
with ground level E(1,1)= 92.236~K.
\end{table*}

\section{PDR models of CH}
\label{Appendix-PDR-model}

As in previous papers by the PDRs4All team, we modeled the PDR component of \mbox{d203-506} using the Meudon PDR code \citep{Petit06}. We analyzed the same output model  presented
by \citet{Berne_ch3p} to interpret the presence of \ch{CH3+} in this disk.
This models adopts $G_0$\,=\,2$\times$10$^4$ and 
$n_{\rm H}$\,=\,10$^7$\,cm$^{-3}$. The specific hydrocarbon chemistry
is presented in \citet{goicoechea2025pdrs4all}.
Figure~\ref{fig:CH_models} (upper panel) shows the basic physical structure of the PDR component. 
The {middle panel} shows the C$^+$/C/CO transition zone and the predicted \ch{CH+}, \ch{CH3+}, and \ch{CH} abundance profiles (with respect to H nuclei). 
The {lower panel shows the accumulated column densities of \ch{CH+}, \ch{CH3+}, and \ch{CH} as a function of depth into the disk PDR.}

{Close to the hot PDR surface, where \mbox{$n$(H)\,$\ge$\,$n$(H$_2$)}, we predict that CH formation is dominated by reactions of atomic C($^3P$) with \mbox{FUV-pumped}, vibrationally excited H$_2$($v$), which is a very endoergic reaction\footnote{We model this reaction from 
\citet{Dean91} by adopting 
\mbox{$v$-state-dependent} rate constants where the energy $E_{v,J}$ of each H$_2$($v$) ro-vibrational state is
subtracted from the reaction endoergicity $\Delta E$ 
(when \mbox{$\Delta E > E_{v,J}$}). That is, 
\mbox{$k_{v,J}(T)$\,$\propto$\,exp$(-[\Delta E - E_{v,J}]/k_{\rm B}T)$}.}, 
about $\sim$11,000\,K when $v$\,=\,0 \mbox{\citep[e.g.,][]{Dean91}}. This hot chemistry leads
to the first CH abundance peak in \mbox{Fig.~\ref{fig:CH_models}}.
In addition,  reactions of  electronically excited neutral carbon atoms \citep[e.g., C($^1S$) and C($^1D$),  as detected by ][]{Haworth23,Goicoechea2024} with H$_2$ likely contribute to
enhance the production of CH. 
These reactions are highly exothermic \mbox{\citep[e.g.,][]{Lezana07}}, but they are currently not included in our models or in other irradiated disk models.
Still, we expect that most of the CH column density arises from deeper layers within the molecular PDR, beyond the \ch{CH3+} abundance peak. In these regions, CH formation is driven by the dissociative recombination of \ch{CH3+} ions, which either directly produce CH or yield \ch{CH2} molecules that subsequently react with H atoms to form CH.}

This photochemical model predicts a total CH
column density $N$(CH)\,$\lesssim$10$^{14}$\,cm$^{-2}$ across the PDR,
and gas temperatures varying from $\sim$4000\,K at the rim of the PDR to
several hundred in the more FUV-shielded layers. These temperatures are broadly consistent with the excitation temperature derived using \mbox{PAHTATmol}.
{Still, non-LTE excitation models will be required to accurately determine the true IR-emitting CH column density and its precise location within the disk PDR. }

\begin{figure}[h!]
\centering
\includegraphics[width=0.5\textwidth]{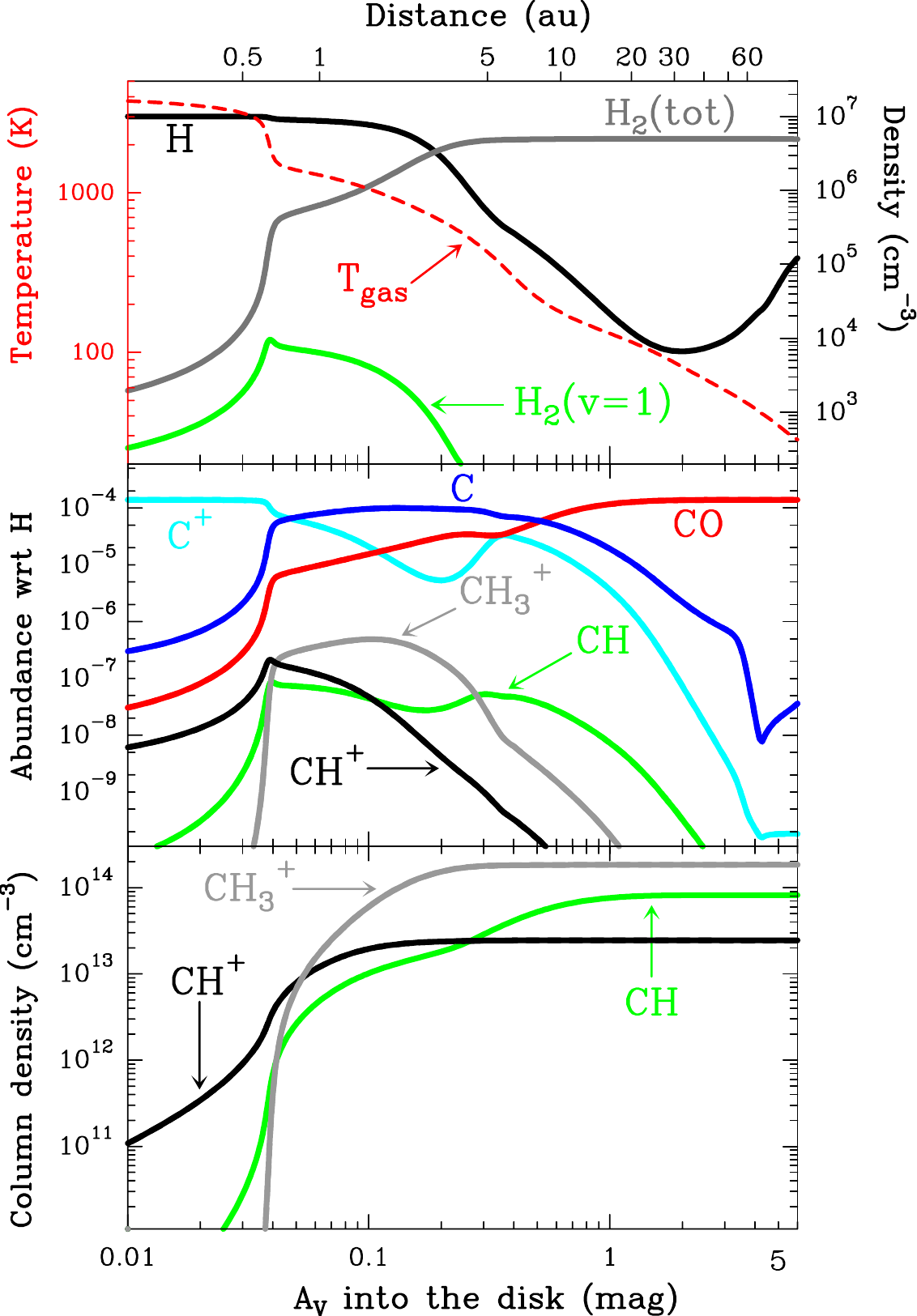}
\caption{Predicted physical (roughly vertical) structure of the PDR component of d203-506. 
\textit{Upper panel}: Gas density and temperature  profiles as a function of depth into the PDR. 
\textit{Middle panel}: Abundance profiles of selected species. \textit{Lower panel}: Column densities as a function of depth into the PDR of small hydrocarbons detected by JWST in \mbox{d203-506} \citep[see also][]{BerneO_2024_science,zannese2025}.}
\label{fig:CH_models}
\end{figure}

\end{document}